\documentstyle[11pt,newpasp,twoside]{article}
\input epsf

\newcommand{\EQ}{\begin{equation}}
\newcommand{\EN}{\end{equation}}
\newcommand{\EQA}{\begin{eqnarray}}
\newcommand{\ENA}{\end{eqnarray}}
\newcommand{\eq}[1]{(\ref{#1})}

\newcommand{\Eq}[1]{equation~(\ref{#1})}

\newcommand{\Fig}[1]{Fig.~\ref{#1}}

\newcommand{\bra}[1]{\langle #1\rangle}

{}

\newcommand{\meanBB}{\overline{\bf{B}}}
\newcommand{\meanJJ}{\overline{\bf{J}}}

{}
{}
{}
{}
{}
{}
{}

\newcommand{\uu}{{\bf{u}}}
\newcommand{\BB}{{\bf{B}}}

\newcommand{\JJ}{{\bf{J}}}
\newcommand{\jj}{{\bf{j}}}

\newcommand{\bb}{{\bf{b}}}

\newcommand{\nab}{\mbox{\boldmath $\nabla$} {}}
\newcommand{\OO}{\mbox{\boldmath $\Omega$} {}}

\newcommand{\yapj}[3]{ #1, {ApJ,} {#2}, #3}
\newcommand{\yapjl}[3]{ #1, {ApJ,} {#2}, #3}

\newcommand{\yan}[3]{ #1, {AN,} {#2}, #3}

\newcommand{\yana}[3]{ #1, {A\&A,} {#2}, #3}

\newcommand{\yjfm}[3]{ #1, {JFM,} {#2}, #3}
\newcommand{\ypf}[3]{ #1, {Phys. Fluids,} {#2}, #3}

\newcommand{\yprl}[3]{ #1, {PRL,} {#2}, #3}

\newcommand{\ybook}[3]{ #1, {#2} (#3)}

\def\edcomment#1{\iffalse\marginpar{\raggedright\sl#1\/}\else\relax\fi}
\reversemarginpar

\begin{document}
\title{Ejection of bi-helical fields from the sun}

\author{Axel Brandenburg}
\affil{Nordita, Blegdamsvej 17, DK-2100 Copenhagen \O, Denmark}

\author{Eric G.\ Blackman}
\affil{Department of Physics \& Astronomy, University of Rochester,
Rochester NY 14627, USA}

\begin{abstract}
It is argued that much of the observed magnetic helicity losses at the
solar surface may represent a reduction of an otherwise more dominant
nonlinearity of solar and stellar dynamos.
This nonlinearity is proportional to the internal twist (as opposed
to writhe) of helical and sigmoidal surface structures.
\end{abstract}


\section{Introduction}

Dynamo action is possible both for helical and nonhelical turbulence
[see, e.g., Meneguzzi et al.\ (1980) for early numerical work].
However, the reason one is particularly interested in helical dynamos is
that there is a well-known mechanism (the $\alpha\Omega$ mechanism)
driving cyclic behavior together with latitudinal migration of a large
scale field.
Both cyclic behavior and latitudinal migration are important
features of dynamos in the sun and many late-type stars.

The $\alpha$-effect is formally introduced via the longitudinally averaged
induction equation.
This effect characterizes the conversion of toroidal mean field
into poloidal via a sequence of events, all of which produce
similarly oriented loops that are tilted in a clockwise sense against
the toroidal direction in the northern hemisphere, and anticlockwise in
the southern.
This effect is the result of the combined action of Coriolis force and
nonuniformity of the turbulence, as embodied by the formula
(e.g.\ R\"udiger \& Kitchatinov 1993)
\EQ
\alpha_0\approx-\tau^2\overline{\uu^2}\,
\OO\cdot\nab\ln(\rho\overline{\uu^2}).
\label{standardalpha}
\EN
This equation, which has been obtained using different 
approaches (see Moffatt 1980), reflects the fact that the turbulent
velocity field has attained {\it kinetic} helicity.
The subscript 0 indicates that the nonlinear feedback is not included.

\section{Connection with magnetic helicity}

It is first of all important to realize that practically no net magnetic
helicity can be generated in the sun.
What is possible, however, is a {\it segregation} of the magnetic field
into its positively and negatively helical constituents.
Such a segregation can occur either in wavenumber space or in real space.
The latter can be accomplished by differential rotation acting on field
lines crossing the equator, while the former can be the result of the
$\alpha$-effect.
This will now be discussed in more detail.

\begin{figure}[t]
\begin{center}\epsfbox{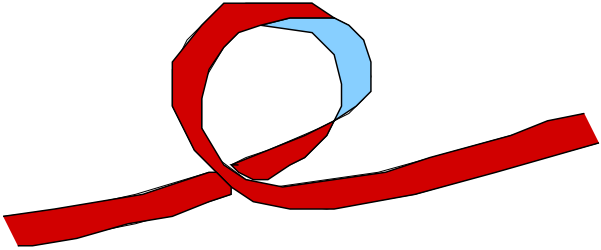}\end{center}\caption[]{
Tilting of the rising tube due to the Coriolis force.
Note that the tilting of the rising loop causes also internal twist.
}\label{ribbon2}\end{figure}

Consider the rise of a toroidal flux tube lifted upward either by
thermal or magnetic buoyancy.
As it rises, it gets tilted by the Coriolis force and, as a consequence,
it must develop some internal twist.
This is already quite clear from a simple sketch (\Fig{ribbon2}),
where the flux tube is depicted as a ribbon, so one can more easily
trace the induced twist.

\epsfxsize=8cm

Mathematically, magnetic helicity is the sum of writhe
and twist helicities, and their sum must stay nearly unchanged
(magnetic helicity conservation).
In the example shown in \Fig{ribbon2}, the overall structure of the tube
follows the right hand rule, corresponding to positive writhe helicity.
At the same time, the interval twist of the tube follows the
left hand rule, corresponding to negative twist helicity.
On the southern hemisphere both signs would be reversed.

\section{Writhe and twist as driver and quencher in dynamo theory}

We have already eluded to the fact that the tilted magnetic field from
the $\alpha$-effect contributes directly (via many systematic and similar
events) to the large scale poloidal field, $\meanBB_{\rm pol}$.
Since $\alpha$ is positive, it should give a positive contribution
to the current helicity of the mean field, i.e.\ $\meanJJ\cdot\meanBB>0$.
We also know from helicity spectra of magnetic flux tube experiments
that the negative internal twist contributes
at scales smaller than the positive writhe
(Blackman \& Brandenburg 2003).
To put this into numbers, we expect the scales of the large scale poloidal
field to be on the order of around 300 Mm, which corresponds to the
latitudinal extent of the toroidal flux belts of around 20-30$^\circ$.
The typical scale associated with the twist is expected to be smaller,
perhaps 30-300 Mm, but the dividing line between small and large scales
may not be very clear.
(Obviously, what is small scale to a dynamo theorist may be large scale
to a solar observer!)

The internal twist is of tremendous importance in dynamo theory.
As explained in the previous section, if the writhe helicity is composed
of large scale field, the twist helicity must be composed of small scale
field (where `small' could be anywhere between 30-300 Mm!).
We may therefore identify the helicity from the internal twist with the
current helicity of the small scale field, $\overline{\jj\cdot\bb}$.
Here, lower characters denote the small scale field, i.e.\ we have
decomposed the field as $\BB=\meanBB+\bb$ and likewise the current as
$\JJ=\meanJJ+\jj$.

The significance of the $\overline{\jj\cdot\bb}$ term is that it gives
an extra contribution to the $\alpha$-effect (Pouquet et al.\ 1976),
\EQ
\alpha=\alpha_0+{\textstyle{1\over3}}\tau\overline{\jj\cdot\bb}/\rho,
\label{PFLformula}
\EN
but it acts such as to suppress or quench the dynamo ($\alpha_0$ and
$\overline{\jj\cdot\bb}$ tend to have opposite signs).
Indeed, the $\overline{\jj\cdot\bb}$ term constitutes the main
nonlinearity of $\alpha$-effect dynamos.
There are other nonlinearities, such as a direct suppression of terms in
\Eq{standardalpha}, but such nonlinearities are never catastrophic in the
sense that they do not depend on the magnetic Reynolds number\footnote{
By contrast, \Eq{PFLformula} can lead to a
catastrophic quenching formula:
use Keinigs' (1983) formula for the {\it saturated} state,
$\alpha=-\eta\overline{\jj\cdot\bb}/\meanBB^2$ (which follows
from the magnetic helicity equation), eliminate
$\overline{\jj\cdot\bb}$ from \eq{PFLformula}, to get
catastrophic quenching:
$\alpha=\alpha_0/(1+R_{\rm m}\meanBB^2/B_{\rm eq}^2)$.
Here we have used $R_{\rm m}=\eta_{\rm t}/\eta$, with
$\eta_{\rm t}={1\over3}\tau\overline{\uu^2}$ and defined
$B_{\rm eq}=(\mu_0\rho\overline{\uu^2})^{1/2}$ to eliminate $\tau$;
see Eqs~(18) and (40) of Blackman \& Brandenburg (2002) for a more
general formulation.}.

\section{Why coronal mass ejections might be good for the dynamo}

Active regions and coronal mass ejections are the main contributors of
magnetic helicity flux from the solar surface
(e.g.\ D\'emoulin et al.\ 2002).
It might not be appropriate to associate them with small scale losses
only.
Instead, in view of the combined presence of large and small scale fields
in a single field structure (\Fig{ribbon2}), such losses occur probably
simultaneously at large and small scale fields.
In this section we argue that this leads to an optimal scenario for
dynamos.

Phenomenologically, surface losses of large and small scale helical fields
may be described by diffusion terms.
This concept has been tested against simulations in the case where losses
occur almost entirely at large scales (Brandenburg \& Dobler 2001).
The results are twofold: the saturation time is formally decreased,
but only at the expense of lowering the saturation field strength.

\epsfxsize=8.0cm\begin{figure}[t!]
\begin{center}\epsfbox{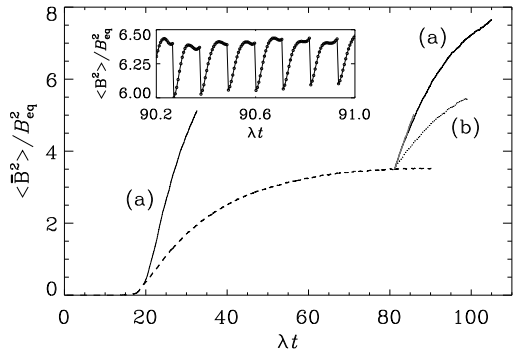}\end{center}\caption[]{
The effect of removing small scale magnetic energy in regular time
intervals $\Delta t$ on the evolution of the large scale field (solid
lines).
The dashed line gives the evolution of $\bra{\meanBB^2}$ for Run~3 of
Brandenburg (2001), where no such energy removal was included.
The two solid lines show the evolution after
restarting at $\lambda t=20$ and $\lambda t=80$.
Time is scaled with the kinematic growth rate $\lambda$.
The curves labeled (a) give the result for $\Delta t=0.12\lambda^{-1}$
and those labeled (b) for $\Delta t=0.4\lambda^{-1}$.
The inset shows, for a short time interval, the sudden drop and subsequent
recovery of the total (small and large scale) magnetic energy in regular
time intervals (adapted from Brandenburg et al.\ 2002).
}\label{pbmean}\end{figure}

The rest of a possible success story is still speculation,
but is based on physical reasoning and a numerical experiment:
if losses occur preferentially on small scales, then the saturation
amplitude of the large scale field is increased.
This is because $\overline{\jj\cdot\bb}$, the `quencher' in dynamo
theory, is constantly being removed. This is confirmed by simulation
where small scale magnetic field is artificially removed (\Fig{pbmean}).
Even in the absence of such losses, the saturation
value of the large scale field may be sufficient for the
sun, but the cycle period may be too long. The primary importance
of small scale losses may actually be to maintain the observed cycle
period. This needs further investigation.

\section{Conclusions}

An important missing link in the story outlined above is, in our
view, a simulation of a coronal mass ejection and an analysis in
terms of the magnetic helicity budget.
At the same time, it will be necessary to improve mean field theory
to take helicity fluxes properly into account.
For attempts in that direction we refer to recent work
by Kleeorin and collaborators (2003).


\begin{references}

\reference
Blackman, E.\ G., \& Brandenburg, A.\yapj{2002}{579}{359}

\reference
Blackman, E.\ G., \& Brandenburg, A.\yapjl{2003}{584}{L99}

\reference
Brandenburg, A.\yapj{2001}{550}{824}

\reference
Brandenburg, A., \& Dobler, W.\yana{2001}{369}{329}

\reference
Brandenburg, A., Dobler, W., \& Subramanian, K.\yan{2002}{323}{99}

\reference
D\'emoulin, P., Mandrini, C.\ H., van Driel-Gesztelyi, L.\yapj{2002}{382}{650}

\reference
Keinigs, R. K.\ypf{1983}{26}{2558}

\reference
Kleeorin, N. I, Kuzanyan, K., Moss, D., et al.\yana{2003}{409}{1097}

\reference
Meneguzzi, M., Frisch, U., \& Pouquet, A.\yprl{1981}{47}{1060}

\reference
Moffatt, H. K.\ybook{1978}
{Magnetic Field Generation}
{Cambridge University Press}

\reference
Pouquet, A., Frisch, U., \& L\'eorat, J.\yjfm{1976}{77}{321}

\reference
R\"udiger, G., \& Kitchatinov, L. L.\yana{1993}{269}{581}

\end{references}
\end{document}